**Imaging the noncentrosymmetric structural organisation of tissue with Interferometric Second Harmonic Generation microscopy**

Maxime Rivard[1], Konstantin Popov[2], Mathieu Laliberté[1], Antony Bertrand-Grenier[1], François Martin[1], Henri Pépin[1], Christian P. Pfeffer[3], Cameron Brown[4], Lora Rammuno[2], François Légaré[1]

[1] Institut National de la Recherche Scientifique, Centre Énergie Matériaux Télécommunications, 1650 Boulevard Lionel-Boulet, Varennes, Qc Canada J3X1S2

[2] University of Ottawa, Department of Physics, MacDonald Hall, 150 Louis Pasteur, Ottawa, On Canada K1N 6N5

[3] Department of Heart Surgery, Ludwig-Maximilians-University Munich, Marchioninistrasse 15, 81377 Munich

[4] Botnar Research Centre, NDORMS, University of Oxford, UK

We report the imaging of tendon, a connective tissue rich in collagen type I proteins, with Interferometric Second Harmonic Generation (I-SHG) microscopy. We observed that the noncentrosymmetric structural organization can be maintained along the fibrillar axis over more than 150 μm, while in the transverse direction it is ~1-15 μm. Those results are explained by modeling tendon as a heterogeneous distribution of noncentrosymmetric nano-cylinders (collagen fibrils) oriented along the fibrillar axis. The preservation of the noncentrosymmetric structural organization over multiple tens of microns reveals that tendon is made of domains in which the fraction occupied by fibrils oriented in one direction is larger than in the other.

Second Harmonic Generation (SHG) microscopy is a powerful technique for biomedical imaging. Specifically, it has been used to image connective tissues rich in collagen type I/III [1,2], cartilage tissues rich in collagen type II [3,4], the myosin bands of muscle [5,6], and microtubules [7]. One common aspect of all those biological structures is that they are composed of noncentrosymmetric proteins where, even at spatial scale much larger than the molecule, the lack of center of inversion remains macroscopically.

Despite the important applications of SHG microscopy in biomedical science, this technique cannot provide information about the relative orientation between the noncentrosymmetric structures that compose the tissues since the SHG phase is not measured. To circumvent this limitation, I-SHG microscopy has been used for imaging noncentrosymmetric materials including organic [8] and periodically poled [9] crystals. In this paper, we show that this technique enables us to gain important information about the noncentrosymmetric structural organization of tissues. By imaging tendon, we show that an overall orientation of the noncentrosymmetric structures persists at macroscopic scale, and mostly along the fibrillar axis of the tissue (~150 μm). Using a numerical model, we show that the preservation of an overall noncentrosymmetricity in the tissue arises because on average more fibrils are pointing along one direction within macroscopic domains.

The experimental setup used for I-SHG microscopy is described in the supplemental material [10]. In I-SHG, images are acquired as a function of the relative phase between second harmonic generated outside the microscope and in the sample. This relative phase is named the reference beam phase (RBP). The raw images acquired with this setup contain signals that are independent of the RBP ($I_{ref}$ and $I_{exp}$). To isolate the interferometric contrast, two images with a RBP difference of 180° are subtracted. As shown in equations 1 and 2, this operation not only removes $I_{ref}$ and $I_{exp}$, but also increases the interferometric contrast by a factor of two. Signal above zero in the calculated image is shown on a green scale while signal below zero is shown on a red scale allowing the imaging of $\chi^{(2)}$ domains. However, since the exact value of the SHG phase in the sample $\varphi_{exp}$ is unknown, it is necessary to take many pairs of images at different RBP to maximize the interferometric contrast.

$$I_{ref+\theta} = I_{ref} + I_{exp} + 2\sqrt{I_{ref}I_{exp}}\cos(\varphi_{exp} - (\varphi_{ref} + \theta)) \quad (1)$$

$$I_{ref} - I_{ref+\pi} = 4\sqrt{I_{ref}I_{exp}}\cos(\varphi_{exp} - \varphi_{ref}) \quad (2)$$

To demonstrate that the I-SHG microscope was working, it was first tested with a periodically poled lithium niobate (PPLN) crystal. As observed in figure 1(a), the $\chi^{(2)}$ domains can already be seen in the SHG image, but I-SHG microscopy enables to see directly the relative phase between them. Figure 1(b) corresponds to the average signal measured in a Region Of Interest (ROI) of an unpoled area (delimited in yellow at the top right of figure 1(c)) of the calculated images for different RBP. As expected, the sinusoidal shape corresponds to the isolated interferometric contrast given by equation 2. In the calculated images for which contrast is maximum (figure 1(c,d)), the poled areas appear in green while the rest of the crystal is shown in red for $\varphi_{ref} = 90°$ while it is the opposite for $\varphi_{ref} = 270°$.

A thin slice (10 μm thick) of tendon from a mouse tail was imaged with standard SHG microscopy to locate an area of interest and for comparison with the interferometric images. As in other connective tissues, collagen fibrils are aligned together within sheets and while SHG microscopy does not provide enough resolution to see them individually, long aligned structures appear in the image (figure 2(a)). The SHG signal changes slowly along the fibrillar axis, but changes on a very short scale in the transverse direction [11].

Interferometric images of the same area reveal more about the organisation of the $\chi^{(2)}$ structures formed by the collagen fibrils at the microscopic scale. Fibrils are grouped together and form long and relatively wide $\chi^{(2)}$ domains, distinguishable by their red or green color, that generate what appears to be π phase shifted SHG signal (figure 2(c,d)). Just like in PPLN, the average interferometric contrast measured in a domain as a function of the RBP has a sinusoidal behaviour (figure 2(b), obtained from the average of the yellow ROI in figure 2(c)).

Two interferometric images with a relative phase of 180° as presented in figure 2(c,d) are not enough to understand how the phase of the signal evolves in tendon within the structures observed with standard SHG microscopy. While individual fibrils have $\pi$ phase shifted piezoelectric tensors at the nanoscale level [12,13], the microscopic $\chi^{(2)}$ domains are formed by a large number of fibrils and the relative phase of their SHG should not be as clearly defined as for PPLN. For this reason, it is necessary to scan the RBP to find where the interferometric contrast is maximized.

The relative phase between individual domains can be found by plotting the interferometric contrast as a function of the RBP and of the position along profiles in the images. In a longitudinal profile along the fibrillar axis as the one in figure 3(b), the SHG phase in tendon can be preserved along very important distance of over 150 μm. This is comparable to the length of an individual collagen fibril [14]. The sinusoidal contrast curves forming the surface would be shifted along the RBP axis if the SHG phase was not maintained along the whole length of the longitudinal profile. In a transverse profile however, the relative phase between the $\chi^{(2)}$ domains can change very rapidly over distance between 1 to 15 μm. Figure 3(c) presents a histogram of the distribution of RBP at which interferometric contrast is maximized within different $\chi^{(2)}$ domains selected along the six transverse profiles in figure 3(a). This histogram reveals that the majority of the domains generate, on average, SHG around two central phases, near 102° (green Gaussian, standard deviation $\sigma = 31°$) and 264° (red Gaussian, $\sigma = 38°$). The 162° difference between these two peaks is close to 180° which corresponds to what would be observed at the nanoscale in individual fibrils. Those results suggest that tendon is similar to PPLN, but the relative phase between its $\chi^{(2)}$ domains is not as well defined.

To facilitate our interpretation of the results, we have performed numerical simulations of the SHG in fibrillous samples. We represent the sample by a collection of infinitely long cylinders of diameter $D_f$ aligned perpendicular to the laser propagation axis. All the cylindrical axes are parallel to each other and to the laser polarization and cross the plane that is perpendicular to the axes at randomly chosen points. The spatial distribution of these points is uniform, and the entire model tissue is characterized by the volume filling ratio $\rho$.

Every cylinder is associated with a scalar nonlinear susceptibility $\chi^{(2)}$, constant throughout it, but with a sign that can be either positive or negative. It is assumed that the total fraction $f$ of all the cylinders have positive sign of their associated $\chi^{(2)}$ and the fraction $(1 - f)$ have the negative sign of $\chi^{(2)}$. This model structure is supported by electron microscopy and piezoelectric force microscopy (PFM) images of connectives tissues rich in collagen type I [12,13]. In particular, PFM has revealed that the noncentrosymmetric structural organization is maintained over the length of the collagen fibrils, but can change direction between adjacent fibrils within the tissue [13]. A similar model was used to simulate the anisotropy of the SHG signal in tendon [15].

The model tissue interacts with a focused continuous wave 1064 nm laser described by a Gaussian beam of width $w_0 = 1.1$ μm. To simplify simulations, no refraction index mismatch in the model tissue was assumed (n=1.4). The SHG signal arises from the coherent sum of the signals generated in all the cylinders. The signal generated by the individual fibrils as measured at the laser axis in the far field zone is calculated numerically using a vectorial modification of the standard scalar Green's function approach [16]. In this approach, the laser-excited nonlinear polarization is multiplied by the Green's function and integrated throughout the cylinder volume to evaluate the amplitude and phase of the SHG signal at a point in the far field zone.

In the context of interferometry, the main quantity of interest is the signal phase. In the results below, the phase $\varphi$ is defined such that the SHG signal measured in the far-field zone at the laser axis from a very thin cylinder located in the focal plane and associated with a positive sign of $\chi^{(2)}$ has $\varphi = 0$. Since the samples are randomly generated, the phase of the signal is not deterministic: it generally depends on the local random tissue composition. As the laser beam scans the tissue, the local fibril composition changes, thus causing changes in the SHG phase.

The simulated phase as a function of the focus location in the model tissue scanned in the transverse direction (perpendicular to the fibrillar axis and the laser propagation) is shown in figure 4(a). If the fraction $f$ of cylinders with a positive $\chi^{(2)}$ is equal to that with a negative $\chi^{(2)}$ ($f = 1 - f = 0.5$) the signal phase continuously changes in the entire $[-\pi; \pi]$ interval as the sample is scanned, without any evident preferred value (solid line in figure 4(a)). We thus conclude that the interferometric images from tissues composed of fibrils with equal fractions of positive and negative signs of their $\chi^{(2)}$ must have a speckle-like appearance, corresponding to a rapidly changing signal phase as the scan is performed. This is not the case for the results presented above. Thus, the model with $f = 0.5$ cannot adequately describe our experimental data.

If the fraction $f$ is increased to 0.55 (dashed line in figure 4(a)), there are almost always more cylinders with the positive sign of $\chi^{(2)}$ in the focal volume. As a result, the interference in the far-field zone produces a stable phase $\varphi \approx 0$ that is almost independent on the scanning position. On the other hand, if $f = 0.45$ (dash-dotted line in figure 4(a)), the phase locks near $\varphi \approx \pm\pi$. We thus conclude that the experimental data can be qualitatively reproduced if one assumes an unequal fraction of the fibrils with different signs of $\chi^{(2)}$. Additionally, this fraction $f$ must be varying in the tissue over short distance transversally to the fibrils and over very long distance longitudinally to explain the domains observed in figure 2(c,d). Since the phase of the SHG shown in figure 3(b) is preserved over long distances, the average number of fibrils with different signs of $\chi^{(2)}$ in the tissue and their structural arrangement must be preserved over important length longitudinally.

The above statement can be characterized more quantitatively. Since the value of the phase is determined by the random local fibril arrangement, one of the characteristics can be the probability of the phase to change by a large value as the tissue is scanned over a distance *d*. Such a characteristic is shown in figure 4(b), where probability of cos($\varphi$) to change by 1 during the scan of $d = 10$ μm is plotted. For the parameters of simulations, this probability becomes negligible (< 10%) for *f* outside of interval (~0.47; ~0.53).

Another quantitative estimation can be made by measuring the SHG phases from a large number of randomly generated tissues and comparing the result with that in figure 3(c). Three sample phase distributions are given by figure 4(c,d,e). In the case of a perfect balance (*f* = 0.5), the phase is uniformly distributed in the entire interval while in an unbalanced case the distribution becomes a relatively narrowly-peaked Gaussian near the zero phase (*f* = 0.55) or near the ±π phase (*f* = 0.45). The standard deviation of the phase distribution is shown in figure 4(f). Given our experimental standard deviation of $<\delta\varphi^2>^{1/2} \sim 30° - 40°$, we can estimate the tissue imbalance to be of the order ~±0.03. Since a combination of the theoretical phase distribution for 10 μm transverse profiles of tissue with *f* = 0.47 or 0.53 would contain two peaks 180° apart, it appears clear that, in a large tissue, *f* changes transversally between 0.47 and 0.53 on average which explains the observation of domains and matches the theoretical phase distribution with the one observed experimentally in figure 3(c).

In conclusion, I-SHG microscopy reveals that the heterogeneous distribution of noncentrosymmetric fibrils in tendon form $\chi^{(2)}$ domains at the microscopic scale. This is because, locally, the fraction of fibrils pointing in one direction can be larger than in the other. These $\chi^{(2)}$ domains are 1-15 μm wide and can be over 150 μm long. By comparing the experimental SHG phase distribution of multiple domains with the SHG phase distributions obtained with a theoretical model of tendon, it appears that $f$ is varying between 0.47 and 0.53 on average in the tissue rather than remaining constant ($f = 0.5$). The structures observed in standard SHG microscopy images are due to the fluctuations of this fraction at different locations in the tissue. Lastly, I-SHG microscopy can be used to better understand the structural organization of any tissues rich in highly organized noncentrosymmetric proteins and has the potential to shed light on electromechanical cell signalling and the piezoelectric modulation of mechanical behaviour.

**Aknowledgement**

The authors acknowledge the financial support from the following funding agencies; NSERC, FRQNT, MDEIE, CIPI, and ORI. LR acknowledges the financial support of the CRC program and Ontario's MRI. MR acknowledges the financial support of FRQNT for the Ph.D. award, KP for the financial support of Mprime NCE Strategic Postdoctoral Research Program, and CB acknowledges financial support from ARUK.

**Figure captions**

Figure 1: (a) SHG image of the interface of a PPLN crystal (200 X 100 µm). (b) Interferometric contrast of the ROI in (c) in function of the RBP. (c) I-SHG image of the area in (a) obtained when $\varphi_{ref}$ = 90° at which contrast is maximized. (d) As expected, the image obtained when $\varphi_{ref}$ = 270° is the opposite of the previous one.

Figure 2: (a) Forward SHG image of tendon (500 X 150 µm). (b) Interferometric contrast in a ROI in (c) in function of the RBP. (c) I-SHG image of the area in (a) and taken for $\varphi_{ref}$ = 105° at which contrast is maximized for most of the fibrillar structures. (d) Again, the image taken for $\varphi_{ref}$ = 285° is the opposite of the previous one.

Figure 3: (a) Forward SHG image of tendon (500 X 150 µm) with ROI in yellow. (b) Surface plot of the interferometric contrast in function of the RBP and of the position along the thin longitudinal profile drawn in (a). (c) Histogram of the distribution of the RBP for which interferometric contrast is maximized for 196 20 µm long longitudinal domains taken in the six yellow bands in (a). The two Gaussian curve fit correspond to the two peaks in the distribution and were calculated from domains with a RBP between 0° and 180° (for the green curve) and between 180° and 360° (for the red).

Figure 4: Results of numerical simulations for $D_f$ = 150 nm, $\rho$ = 0.75. (a) SHG phase behaviour as the sample is scanned. (b) Probability of $\cos(\varphi)$ to change by 1 or more during a 10 microns-long scan. The dashed line shows a Gaussian fit. (c), (d) and (e) Distribution of SHG signal phases measured from 1000 samples with $f$ = 0.5 (c), $f$ = 0.55 in red (d) and $f$ = 0.45 in green (e). (f) Standard deviation of SHG signal phase distribution as a function of $f$. The dashed line shows an exponential fit.

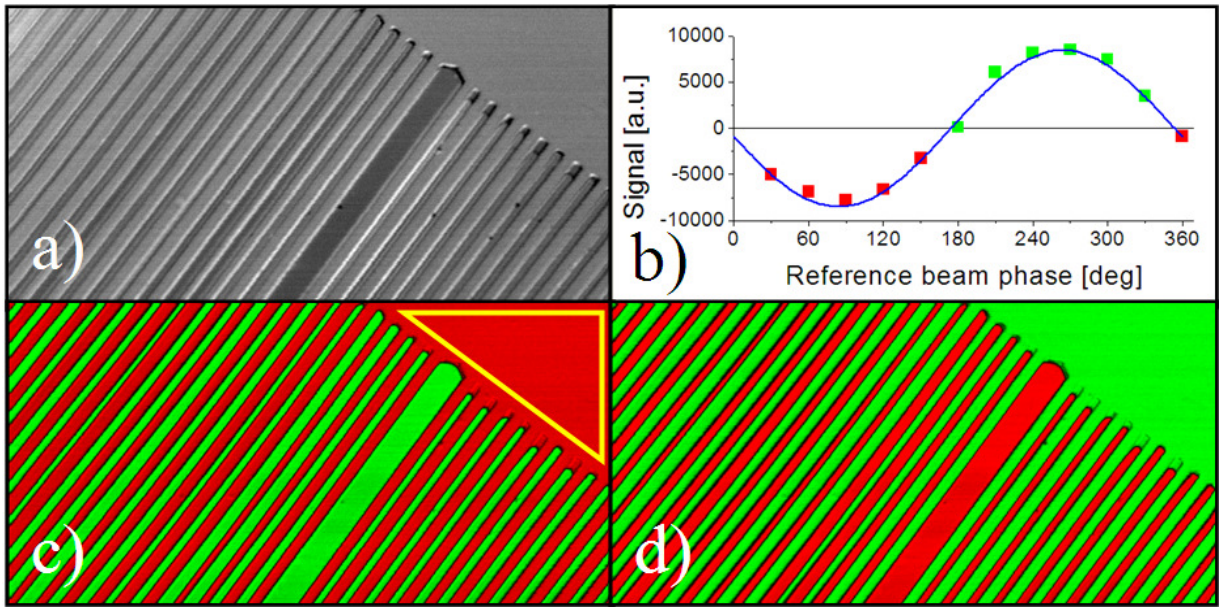

**Figure 1**

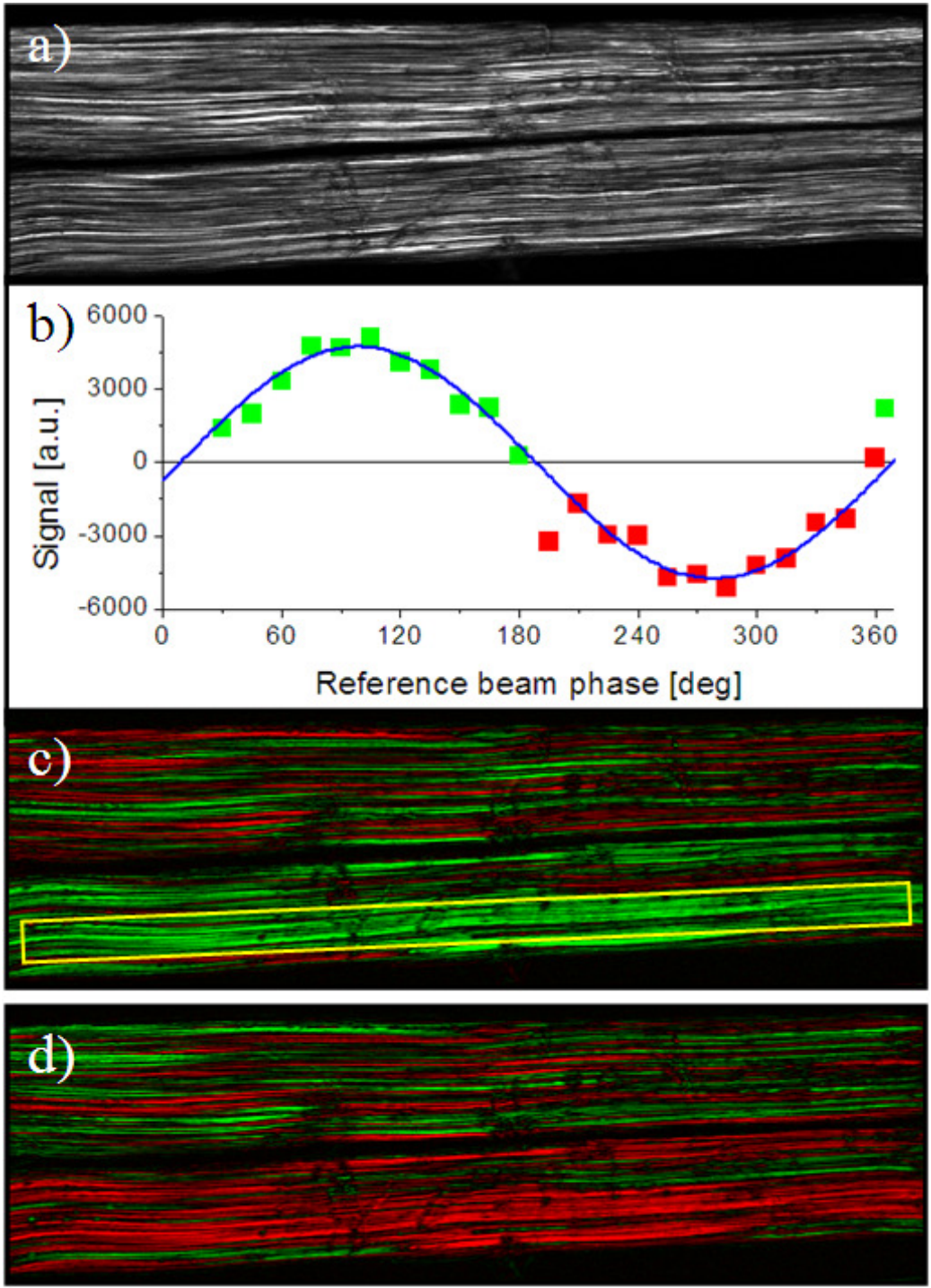

**Figure 2**

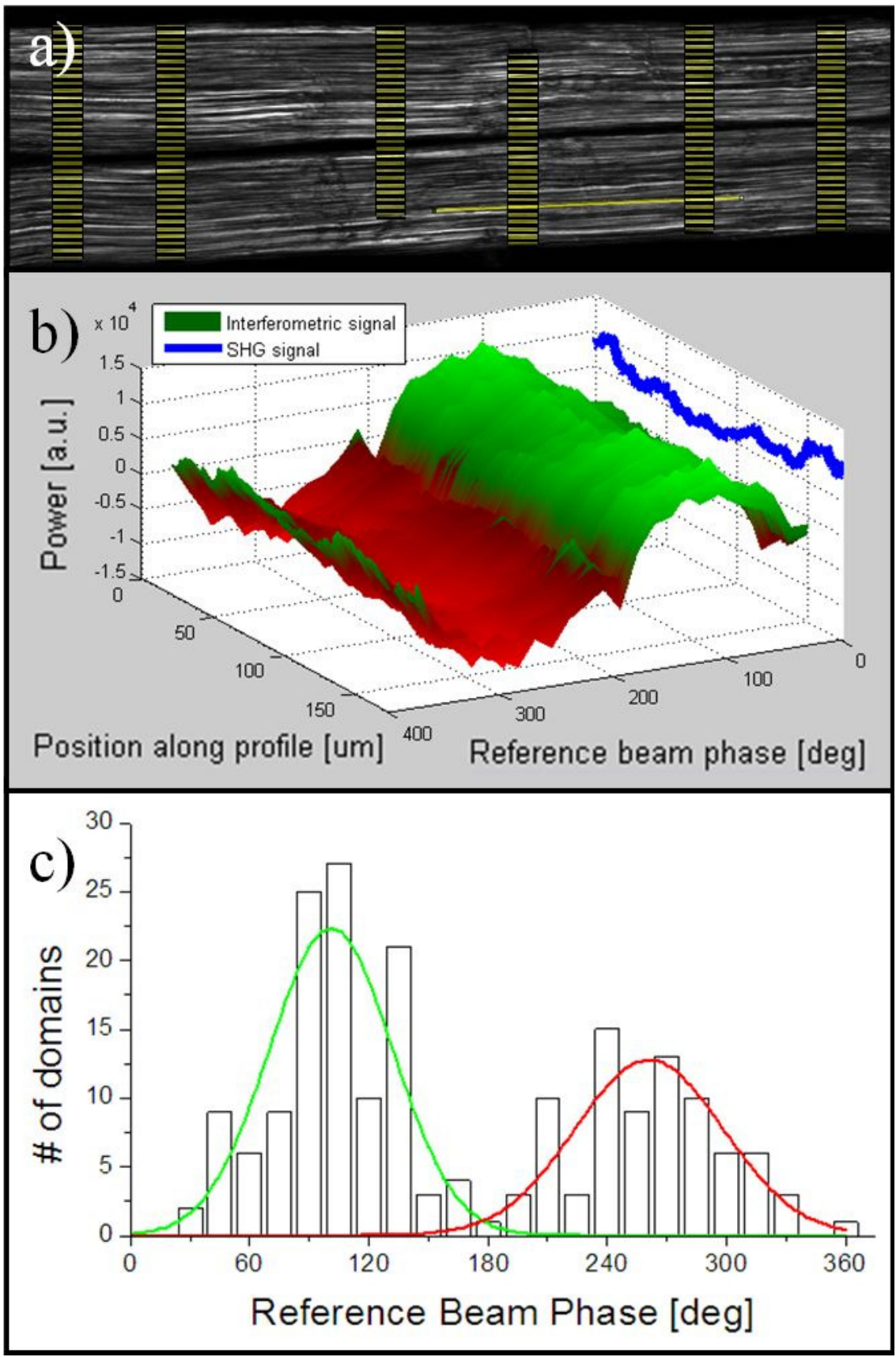

**Figure 3**

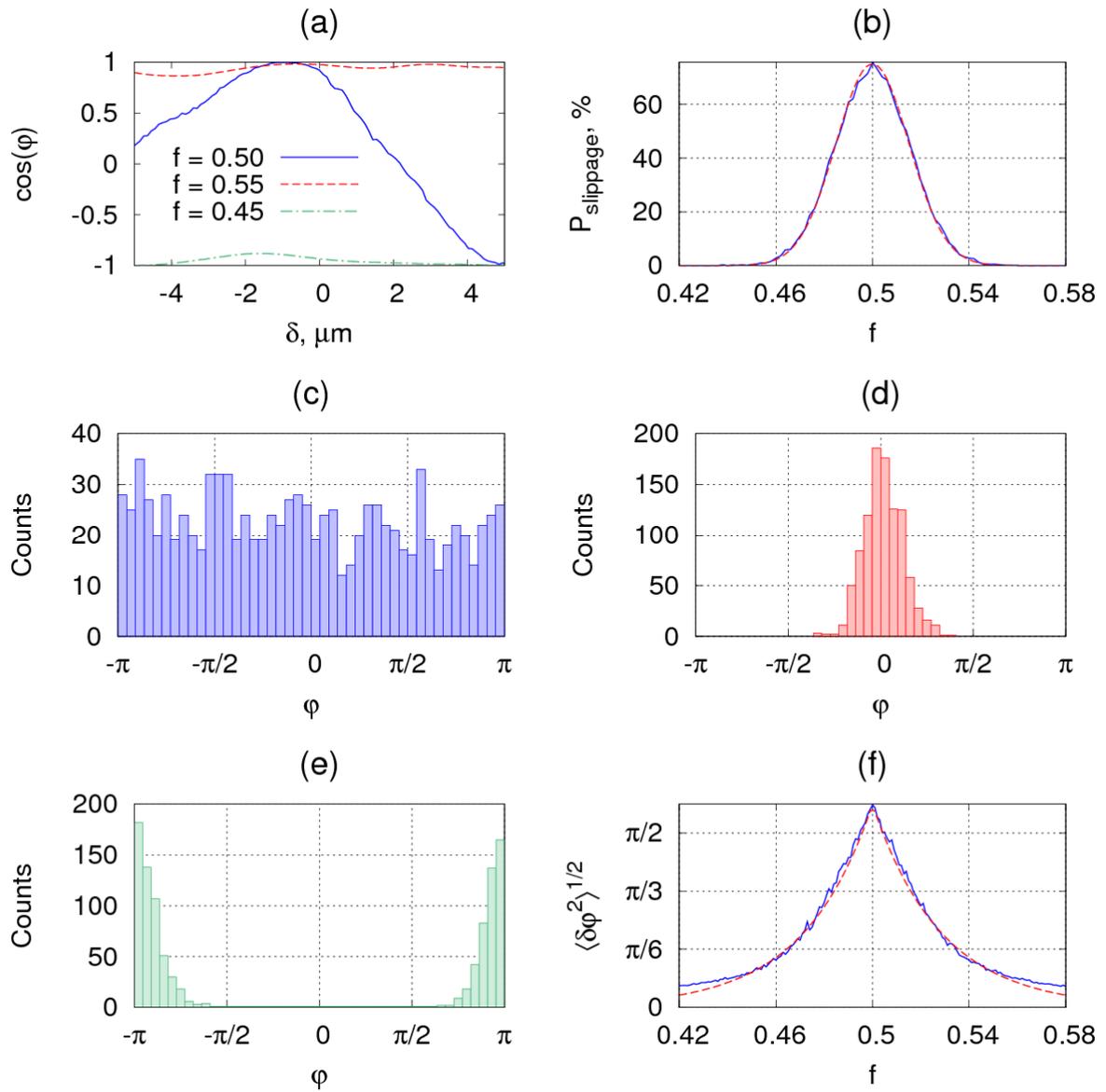

**Figure 4**